\documentstyle{article}

\newcommand{\ket}[1]{| #1 \rangle}

\newdimen \myunit
\newdimen \myhsize
\newdimen \myvsize
\def\sx#1{
\begin{picture}(12,10)(0,0)
\ifcase #1
    \put(0,5){\line(1,0){12}}
\or \put(6,0){\line(0,1){10}}
\or \put(0,5){\line(1,0){12}}
    \put(6,0){\line(0,1){10}}
\fi
\end{picture}
}

\def\dx#1{
\begin{picture}(12,10)(0,0)
  \put(6,5){\circle*{2}}
  \put(0,5){\line(1,0){12}}
\ifcase #1
    \put(6,0){\line(0,1){10}}
\or \put(6,5){\line(0,1){5}}
\or \put(6,5){\line(0,-1){5}}
\fi
\end{picture}
}

\def\nt#1{
\begin{picture}(12,10)(0,0)
  \put(6,5){\circle{4}}
  \put(0,5){\line(1,0){12}}
\ifcase #1
    \put(6,0){\line(0,1){10}}
\or \put(6,3){\line(0,1){7}}
\or \put(6,7){\line(0,-1){7}}
\fi
\end{picture}
}

\def\ox#1{
\begin{picture}(12,10)(0,0)
  \put(6,5){\circle{3}}
  \put(0,5){\line(1,0){12}}
\ifcase #1 
    \put(6,0){\line(0,1){10}}
\or \put(6,10){\line(0,-1){6.5}}
\or \put(6,0){\line(0,1){6.5}}
\fi
\end{picture}
}

\newcommand{\ct}[1]{
\begin{picture}(12,10)(0,0)
  \multiput(0,5)(11,0){2}{\line(1,0){1}}
  \put(6,5){\circle{10}}
  \put(0,0){\vbox to \myvsize{\vfill
	\hbox to \myhsize{\hfill #1\hfill}\vfill}}
\end{picture}
}

\newcommand{\th}[1]{
\begin{picture}(12,10)(0,0)
  \multiput(1,0)(10,0){2}{\line(0,1){10}}
  \multiput(1,0)(0,10){2}{\line(1,0){10}}
  \multiput(0,5)(11,0){2}{\line(1,0){1}}
  \put(0,0){\vbox to \myvsize{\vfill
	\hbox to \myhsize{\hfill #1\hfill}\vfill}}
\end{picture}
}

\newcommand{\place}[1]{\vbox to \myvsize{\vfill
	\hbox to \myhsize{\hfill #1\hfill}\vfill}}
\def\plac#1#2{\vbox to \myvsize{\vfill
	\hbox to #1\myhsize{#2\hfill}\vfill}}
\newcommand{\pv}{\place{\vbox to \myvsize{\vfill\vfill\vfill\vfill\smash{$\vdots$}\vfill}}}
\newcommand{\pcd}{\place{$\cdots$}}

\def\carry#1{
\begin{picture}(24,40)(0,0)
	\multiput(0,5)(0,10){4}{\line(1,0){7}}
	\multiput(17,5)(0,10){4}{\line(1,0){7}}
	\multiput(7,0)(10,0){2}{\line(0,1){40}}
	\multiput(7,0)(0,40){2}{\line(1,0){10}}
	\put(9.5,5){\shortstack{C\\A\\R\\R\\Y}}
\linethickness{1mm}
\ifcase #1
	\put(17,-0.1){\line(0,1){40.3}}
\or \put(7,-0.1){\line(0,1){40.3}}
\fi
\end{picture}}

\def\summy{
\begin{picture}(24,30)(0,0)
	\multiput(0,5)(0,10){3}{\line(1,0){7}}
	\multiput(17,5)(0,10){3}{\line(1,0){7}}
	\put(7,0){\framebox(10,30){}}
	\put(9,6.5){\shortstack{S\\U\\M}}
\end{picture}}

\title{Addition on a Quantum Computer}

\author{Thomas G. Draper}

\date{
Written: September 1, 1998 \hspace{.25in}Revised: June 15, 2000}

\begin{document}
\maketitle

\begin{abstract}
A new method for computing sums on a quantum computer is introduced.  This technique uses the quantum Fourier transform and reduces the number of qubits necessary for addition by removing the need for temporary carry bits.  This approach also allows the addition of a classical number to a quantum superposition without encoding the classical number in the quantum register.  This method also allows for massive parallelization in its execution.
\end{abstract}
\newpage

\tableofcontents
\newpage

\section{Introduction}

Traditionally, addition algorithms designed for a quantum computer have mirrored their classical counterparts\cite{beck,miquel,vedral}, with the necessary extensions for reversible computation.  Faster quantum addition algorithms implement carry-save techniques\cite{gossett,zalka}, but still follow a classical model.  However, the ideal addition algorithm for a quantum computer may not be similar to its classical counterpart.  This paper presents a new paradigm for addition on a quantum computer.

The addition method used takes two values $a$ and $b$, computes $F(a)$ the quantum Fourier transform (QFT) of $a$ and then uses $b$ to evolve $F(a)$ into $F(a+b)$.  The inverse quantum Fourier transform may then be applied and the sum recovered.  Since there is a cost of computing the transform before and after the sum, as much computation as possible should be performed in the transform range before leaving.

This paper assumes a rudimentary background in the ideas of quantum computation.  For an introduction to quantum computation the reader is encouraged to read Art Pittenger's, "An Introduction to Quantum Computing Algorithms"\cite{pitt} or Andrew Steane's, "Quantum Computing"\cite{steane}.  For further study, an excellent searchable pre-print server is maintained by Los Alamos laboratories.  Papers on quantum computing/cryptography may be found at http://xxx.lanl.gov/quant-ph.

\section{Classical Addition}
A number of papers have been published concerning the implementation of addition on a quantum computer \cite{beck,gossett,pitt,vedral,zalka}.  All of the implementations use at least $3n$ qubits to add two $n$-bit numbers.  The method presented here follows the outline in \cite{pitt}.  The adder is composed of two basic unitary computational units.

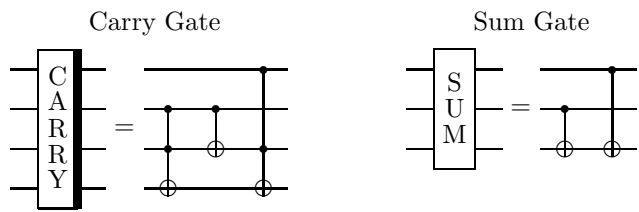
\begin{center}
\begin{picture}(160,60)(0,0)
	\put(22,45){Carry Gate}
	\put(0,0){\carry0}
	\put(25,15){\place{=}}
	\put(30,0)
{\small \renewcommand{\arraystretch}{0}
\begin{tabular}[b]{*{3}{c@{}}}
\sx0&\sx0&\dx2\\
\dx2&\dx2&\sx2\\
\dx0&\nt1&\dx0\\
\nt1&\sx0&\nt1\\
\end{tabular}}

	\put(119,45){Sum Gate}
	\put(100,10){\summy}
	\put(125,20){\place{=}}
	\put(130,10)
{\small \renewcommand{\arraystretch}{0}
\begin{tabular}[b]{*{2}{c@{}}}
\sx0&\dx2\\
\dx2&\sx2\\
\nt1&\nt1\\
\end{tabular}}
\end{picture}
\end{center}

A carry gate with the dark bar on the left side is considered to be a normal carry gate executed in reverse order.  The following diagram shows how these two unitary gates are combined to form a reversible adder.

\begin{center}
{\large Reversible Adder for Two 3-bit Numbers}

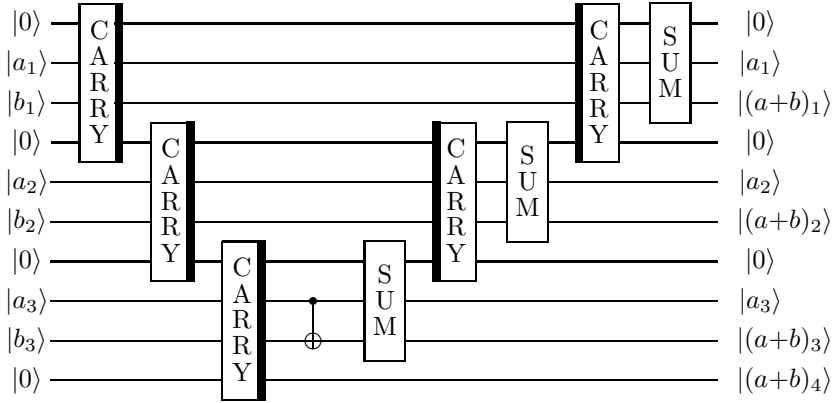
\lineskip.5cm

\begin{picture}(200,100)(-10,0) \put(-10,90){\place{$\ket{0}$}}
									 \put(-10,80){\place{$\ket{a_1}$}}
									 \put(-10,70){\place{$\ket{b_1}$}}
									 \put(-10,60){\place{$\ket{0}$}}
									 \put(-10,50){\place{$\ket{a_2}$}}
									 \put(-10,40){\place{$\ket{b_2}$}}
									 \put(-10,30){\place{$\ket{0}$}}
									 \put(-10,20){\place{$\ket{a_3}$}}
									 \put(-10,10){\place{$\ket{b_3}$}}
									 \put(-10,0){\place{$\ket{0}$}}
									 \put(175,90){\place{$\ket{0}$}}
									 \put(175,80){\place{$\ket{a_1}$}}
									 \put(175,70){\place{$\ket{(a+b)_1}$}}
									 \put(175,60){\place{$\ket{0}$}}
									 \put(175,50){\place{$\ket{a_2}$}}
									 \put(175,40){\place{$\ket{(a+b)_2}$}}
									 \put(175,30){\place{$\ket{0}$}}
									 \put(175,20){\place{$\ket{a_3}$}}
									 \put(175,10){\place{$\ket{(a+b)_3}$}}
									 \put(175,0){\place{$\ket{(a+b)_4}$}}
									 \put(0,60){\carry0}
									 \put(18,30){\carry0}
									 \put(36,0){\carry0}
									 \put(60,30){\sx0}
									 \put(60,20){\dx2}
									 \put(60,10){\nt1}
									 \put(72,10){\summy}
									 \multiput(60,0)(12,0){3}{\sx0}
									 \put(90,30){\carry1}
									 \put(108,40){\summy}
									 \multiput(108,30)(12,0){2}{\sx0}
									 \put(126,60){\carry1}
									 \put(144,70){\summy}
									 \multiput(144,60)(12,0){2}{\sx0}
									 \multiput(2,5)(0,10){3}{\line(1,0){40}}
									 \multiput(2,35)(0,10){3}{\line(1,0){24}}
									 \multiput(18,75)(0,10){3}{\line(1,0){112}}
									 \multiput(42,45)(0,10){3}{\line(1,0){54}}
									 \multiput(96,5)(0,10){3}{\line(1,0){74}}
									 \multiput(132,35)(0,10){3}{\line(1,0){38}}
									 \end{picture}
									 \end{center}

The construction of an n-bit reversible adder is a straightforward extension of the 3-bit adder. Note that an additional $n$ qubits were needed as temporary carry bits.  These qubits are reversibly set back to zero after they are used so they may be used for later computation.  Therefore, even though the input and output may be stored using only $2n$ qubits, $3n$ qubits must be used for the computation.

\section{The Quantum Fourier Transform }
For simplification of notation, let $e(t)=e^{2\pi i t}$.  Let $a\in {\bf Z}_{2^n}$, the additive group of integers modulo $2^n$.  Let $a_{n}a_{n-1}\cdots a_2 a_1$ be the binary representation for $a$, where $a=a_{n}2^{n-1}+a_{n-1}2^{n-2}+\cdots+a_2 2^1+a_1 2^0$.  Then $\ket{a}=\ket{a_{n}}\otimes\ket{a_{n-1}}\otimes\cdots\otimes\ket{a_2}\otimes\ket{a_1}$. The quantum Fourier transform (QFT) of $\ket{a}$ is the mapping
\begin{equation} \label{qft}
\ket{a}\stackrel{F_{2^n}}{\longrightarrow}\frac{1}{2^{\frac{n}{2}}}\sum_{k=0}^{2^n-1}e(ak/2^n)\ket{k}.
\end{equation}
It turns out that (\ref{qft}) is unentangled\cite{cleve}.  Let 
\begin{equation}
\ket{\phi_k(a)}=\frac{1}{\sqrt{2}}(\ket{0}+e(a/2^k)\ket{1}).
\end{equation}
Then (\ref{qft}) factors as
$$\sum_{k=0}^{2^n-1}e(ak/2^n)\ket{k}=\ket{\phi_n(a)}\otimes\cdots\otimes\ket{\phi_2(a)}\otimes\ket{\phi_1(a)}.$$

It is also helpful to notice that $e(a/2^k)=e(0.a_k\ldots a_1)$, where $(0.a_k\ldots a_1)$ is a binary fraction.  Therefore, each $\ket{\phi_k(a)}$ contains the lower $k$ binary digits of $a$.  Consider the following two gate operations:

\[\begin{array}{ccc}
\hbox{Conditional Rotation}&&\hbox{Hadamard Transform}\\ 
{\renewcommand{\arraystretch}{0}
R_k=
\begin{array}{c} \ct{$k$} \\ \dx1 \end{array}}
=
{\renewcommand{\arraystretch}{0}
\begin{array}{c} \dx2 \\ \ct{$k$} \end{array}}
=
\left[ \begin{array}{*{4}{c}} 
1&0&0&0\\ 
0&1&0&0\\
0&0&1&0\\
0&0&0& $$e(\frac{1}{2^k})$$ \end{array} \right]
&\hbox{and}&
\begin{array}{c}\th{H}\end{array}
=
\frac{1}{\sqrt{2}}\left[\begin{array}{cc} 1&1\\ 1&-1 \end{array}\right]
\end{array}
\]

The conditional rotation gate performs a phase rotation between two qubits conditioned on their superposition.  For example, if the state between two qubits was $\alpha\ket{00}+\beta\ket{01}+\gamma\ket{10}+\delta\ket{11}$, then after performing a conditional $k$ rotation gate their joint state would be $\alpha\ket{00}+\beta\ket{01}+\gamma\ket{10}+e(1/2^k)\delta\ket{11}$.  The Hadamard transform gate operates on a single qubit and transforms the state $\alpha\ket{0}+\beta\ket{1}$ to $\frac{1}{\sqrt{2}}((\alpha+\beta)\ket{0}+(\alpha-\beta)\ket{1})$.
 Computing the QFT may be done according to the following wire diagram.

\begin{center}
{\small \renewcommand{\arraystretch}{0}
\begin{tabular}[b]{c@{\hspace*{5\myunit}}*{12}{c@{}}c@{\hspace*{5\myunit}}c}
\multicolumn{15}{c}{{\large Quantum Fourier Transform}}\\
\hspace{15\myunit}&\vspace{5\myunit}&&&&&&&&&&&&&\hspace{20\myunit} \\
\place{$\ket{a_n}$}&
\th{$H$}&\ct{$2$}&\pcd&\ct{$n$-\hskip-\myunit$1$}&\ct{$n$}&
\sx0&\sx0&\sx0&\sx0&\pcd&\sx0&\sx0&\sx0&
{\place{$\ket{\phi_n(a)}$}}\\
\place{$\ket{a_{n-1}}$}&
\sx0&\dx1&\sx0&\sx2&\sx2&
\th{$H$}&\pcd&\ct{$n\hbox{-}2$}&\ct{$n\hbox{-}1$}&\pcd&\sx0&\sx0&\sx0&
{\place{$\ket{\phi_{n-1}(a)}$}}\\
{\pv}&
&&&\sx1&\sx1&&&\sx1&\sx1&\\
\place{$\ket{a_2}$}
&\sx0&\sx0&\sx0&\dx1&\sx2&\sx0&\sx0&\dx1&\sx2&\pcd&\th{$H$}&\ct{$2$}&\sx0&
\place{$\ket{\phi_2(a)}$}\\
{\place{$\ket{a_1}$}}
&\sx0&\sx0&\sx0&\sx0&\dx1&\sx0&\sx0&\sx0&\dx1&\pcd&\sx0&\dx1&\th{$H$}&
{\place{$\ket{\phi_1(a)}$}}\\
\end{tabular}}
\end{center}

To see the effect of the transform on a qubit, we will follow what each gate does to the qubit $\ket{a_n}$.

\begin{center}
\renewcommand{\arraystretch}{1.5}
\begin{tabular}{r@{\hspace{.1cm}$\longrightarrow$\hspace{.1cm}}ll}
$\ket{a_n}$&$\frac{1}{\sqrt{2}}(\ket{0}+e(0.a_n)\ket{1})$&Hadamard transform\\
&$\frac{1}{\sqrt{2}}(\ket{0}+e(0.a_n a_{n-1})\ket{1})$&$R_2$ rotation conditioned on $a_{n-1}$\\
\multicolumn{2}{c}{\vdots}&\hspace{1cm}\vdots\\
&$\frac{1}{\sqrt{2}}(\ket{0}+e(0.a_n a_{n-1}\ldots a_1)\ket{1})$&$R_n$ rotation conditioned on $a_1$\\
\multicolumn{2}{l}{\hspace{1.4cm}$=\ket{\phi_n(a)}$}\\
\end{tabular}
\end{center}

\section{The Approximate QFT}

As $k$ gets large, the conditional rotation associated with $k$ gets very small, and hence, the rotation matrix approaches the identity.  It is natural to ask how good an approximation may be if we do not perform, or are unable to perform, gates below a certain tolerance.  Barenco, et. al. proved that in the presence of decoherence, an approximate quantum Fourier transform (AQFT) may in fact be more accurate than a full QFT\cite{aqft}.  Depending on the decoherence present, the optimal value for $k$ needed is around $\hbox{log}_2 n$.  This reduces the number of operations needed for a quantum Fourier transform from $\frac{1}{2}n(n+1)$ operations to $\frac{1}{2}(2n-\hbox{log}_2n)(\hbox{log}_2n-1)\approx n\hbox{log}_2n$ operations.

\section{Quantum Addition}

The quantum addition is performed using a sequence of conditional rotations which are mutually commutative.  The structure is very similar to the quantum Fourier transform, but the rotations are conditioned on $n$ external bits.  This is helpful if we wish to add classical data to quantum data.  Since all the control bits are known a priori, it is just a matter of programming to implement the resulting rotations.  The addition is performed according to the following diagram.

\begin{center}
{\small \renewcommand{\arraystretch}{0}
\begin{tabular}[b]{c@{\hspace*{15\myunit}}*{12}{c@{}}c@{\hspace*{5\myunit}}c}
\multicolumn{15}{c}{{\large Transform Addition}}\\
\hspace{30\myunit}&\vspace{5\myunit}&&&&&&&&&&&&&\hspace{20\myunit} \\

\place{$\ket{b_n}$}&
\dx2&\sx0&\sx0&\sx0&\sx0&\sx0&\sx0&\sx0&\sx0&\sx0&\sx0&\sx0&\sx0&
\place{$\ket{b_n}$}\\

\place{$\ket{b_{n\hbox{-}1}}$}&
\sx2&\dx2&\sx0&\sx0&\sx0&\dx2&\sx0&\sx0&\sx0&\sx0&\sx0&\sx0&\sx0&
\place{$\ket{b_{n\hbox{-}1}}$}\\

\pv&
\sx1&\sx1&&&&\sx1&&&\\

\place{$\ket{b_2}$}&
\sx2&\sx2&\pcd&\dx2&\sx0&\sx2&\pcd&\dx2&\sx0&
\pcd&\dx2&\sx0&\sx0&\place{$\ket{b_2}$}\\

\place{$\ket{b_1}$}&
\sx2&\sx2&\pcd&\sx2&\dx2&\sx2&\pcd&\sx2&\dx2&
\pcd&\sx2&\dx2&\dx2&\place{$\ket{b_1}$}\\

\place{$\ket{\phi_n(a)}$}&
\ct{1}&\ct{2}&\pcd&\ct{$n$-1}&\ct{$n$}&\sx2&\pcd&\sx2&
\sx2&\pcd&\sx2&\sx2&\sx2&\place{$\ket{\phi_n(a+b)}$}\\

\place{$\ket{\phi_{n\hbox{-}1}(a)}$}&
\sx0&\sx0&\sx0&\sx0&\sx0&\ct{1}&\pcd&\ct{$n$-2}&
\ct{$n$-1}&\pcd&\sx2&\sx2&\sx2&\place{$\ket{\phi_{n\hbox{-}1}(a+b)}$}\\

\pv&
&&&&&&&&&&\sx1&\sx1&\sx1&
\pv\\

\place{$\ket{\phi_2(a)}$}&
\sx0&\sx0&\sx0&\sx0&\sx0&\sx0&\sx0&\sx0&\sx0&\sx0&
\ct{1}&\ct{2}&\sx2&\place{$\ket{\phi_2(a+b)}$}\\

\place{$\ket{\phi_1(a)}$}&
\sx0&\sx0&\sx0&\sx0&\sx0&\sx0&\sx0&\sx0&\sx0&\sx0&
\sx0&\sx0&\ct{1}&\place{$\ket{\phi_1(a+b)}$}\\

\end{tabular}}
\end{center}

Once again it is helpful to trace the state of one qubit during the computation.

\begin{center}
\renewcommand{\arraystretch}{1.5}
\begin{tabular}{r@{\hspace{.1cm}$\longrightarrow$\hspace{.1cm}}ll}
$\ket{\phi_n(a)}$&$\frac{1}{\sqrt{2}}(\ket{0}+e(0.a_n a_{n-1}\ldots a_1+0.b_n)\ket{1})$&$R_1$ rotation from $b_n$\\
&$\frac{1}{\sqrt{2}}(\ket{0}+e(0.a_n a_{n-1}\ldots a_1+0.b_n b_{n-1})\ket{1})$&$R_2$ rotation from $b_{n-1}$\\
\multicolumn{2}{c}{\vdots}&\hspace{1cm}\vdots\\
&$\frac{1}{\sqrt{2}}(\ket{0}+e(0.a_n a_{n-1}\ldots a_1+0.b_n b_{n-1}\ldots b_1)\ket{1})$&$R_n$ rotation from $b_1$\\
\multicolumn{2}{l}{\hspace{1.9cm}$=\ket{\phi_n(a+b)}$}\\
\end{tabular}
\end{center}

Since the implementation of the quantum addition is very similar to the QFT, it is not surprising that there is a more efficient approximate implementation of quantum addition using the same technique as the AQFT.  Using analogous reasoning, it is quite clear that the approximate quantum addition can be computed efficiently using only $n\hbox{log}_2 n$ operations as well.

The critical difference between quantum addition and the QFT is that all of the operations commute with each other in quantum addition, whereas the Hadamard transforms necessary to do the QFT require certain ordering.  It follows that if a quantum computer can implement numerous independent gate operations simultaneously, the run time will decrease proportionally with its capabilities.  Therefore, a quantum computer capable of computing $\frac{n}{2}$ independent 2-qubit gate operations simultaneously, can perform quantum addition in about $n+1$ time slices.  If we are using the AQFT technique of eliminating rotations below a certain threshold, the quantum addition may be performed in $\hbox{log}_2 n$ time slices.  One possible parallel method would be to execute all depth 1 rotations simultaneously, and then all depth 2 rotations, etc.  It is clear that each of these are operating on independent qubits.  In \cite{zalka}, Zalka discusses why we may expect this type of parallelism in a quantum computer.

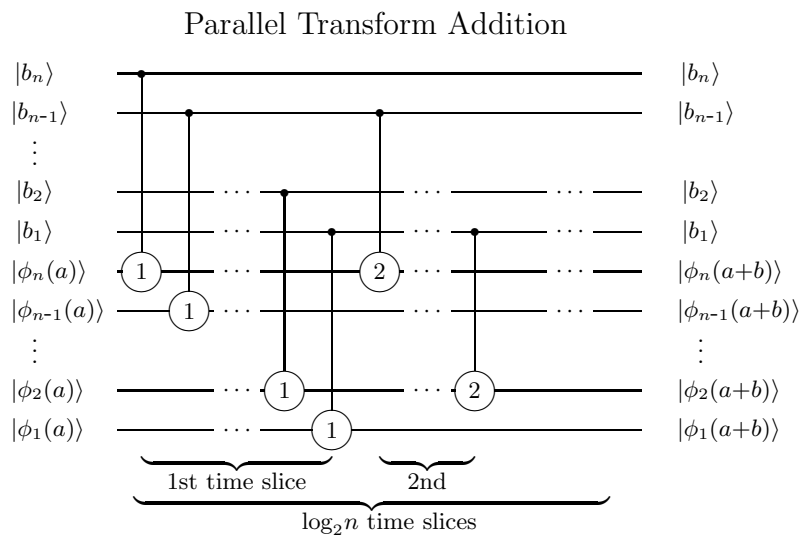
\begin{center}
{\small \renewcommand{\arraystretch}{0}
\begin{tabular}[b]{c@{\hspace*{15\myunit}}*{10}{c@{}}c@{\hspace*{5\myunit}}c}
\multicolumn{13}{c}{{\large Parallel Transform Addition}}\\
\hspace{30\myunit}&\vspace{5\myunit}&&&&&&&&&&&\hspace{20\myunit} \\

\place{$\ket{b_n}$}&
\dx2&\sx0&\sx0&\sx0&\sx0&\sx0&\sx0&\sx0&\sx0&\sx0&\sx0&
\place{$\ket{b_n}$}\\

\place{$\ket{b_{n\hbox{-}1}}$}&
\sx2&\dx2&\sx0&\sx0&\sx0&\dx2&\sx0&\sx0&\sx0&\sx0&\sx0&
\place{$\ket{b_{n\hbox{-}1}}$}\\

\pv&
\sx1&\sx1&&&&\sx1&&&\\

\place{$\ket{b_2}$}&
\sx2&\sx2&\pcd&\dx2&\sx0&\sx2&\pcd&\sx0&\sx0&
\pcd&\sx0&\place{$\ket{b_2}$}\\

\place{$\ket{b_1}$}&
\sx2&\sx2&\pcd&\sx2&\dx2&\sx2&\pcd&\dx2&\sx0&
\pcd&\sx0&\place{$\ket{b_1}$}\\

\place{$\ket{\phi_n(a)}$}&
\ct{1}&\sx2&\pcd&\sx2&\sx2&\ct{2}&\pcd&\sx2&
\sx0&\pcd&\sx0&\place{$\ket{\phi_n(a+b)}$}\\

\place{$\ket{\phi_{n\hbox{-}1}(a)}$}&
\sx0&\ct{1}&\pcd&\sx2&\sx2&\sx0&\pcd&\sx2&
\sx0&\pcd&\sx0&\place{$\ket{\phi_{n\hbox{-}1}(a+b)}$}\\

\pv&
&&&\sx1&\sx1&&&\sx1&&&&
\pv\\

\place{$\ket{\phi_2(a)}$}&
\sx0&\sx0&\pcd&\ct{1}&\sx2&\sx0&\pcd&\ct{2}&\sx0&\sx0&
\sx0&\place{$\ket{\phi_2(a+b)}$}\\

\place{$\ket{\phi_1(a)}$}&
\sx0&\sx0&\pcd&\sx0&\ct{1}&\sx0&\sx0&\sx0&\sx0&\sx0&
\sx0&\place{$\ket{\phi_1(a+b)}$}\\

&\multicolumn{5}{c}{$\underbrace{\hbox to 4\myhsize{\vbox to .1\myvsize{}}}$}&
\multicolumn{3}{c}{$\underbrace{\hbox to 2\myhsize{\vbox to .1\myvsize{}}}$}\\
&\multicolumn{5}{c}{1st time slice}&\multicolumn{3}{c}{2nd}\\
&\multicolumn{11}{l}{$\underbrace{\hbox to 10\myhsize{\vbox to .1\myvsize{}}}$}\\
&\multicolumn{11}{c}{$\hbox{log}_2 n$ time slices}\\
\end{tabular}}
\end{center}

\section{Implications}

In Zalka's paper\cite{zalka}, he outlines a number of methods for computing Shor's factorization algorithm on a quantum computer.  The impact of the quantum addition algorithm has the greatest impact if we are trying to minimize the qubits necessary to perform a computation.  Zalka shows that with a number of refinements, Shor's algorithm may be computed using only $3n$ qubits.  Replacing the addition used by Zalka with quantum addition allows us to perform Shor's algorithm using only $2n$ qubits.  It should be noted that these run times are both $O(n^3)$ in the number of operations.  There exist faster algorithms with a run time of $O(n^2\hbox{log} n)$, but these algorithms require substantially more qubits and the qubit reduction from the quantum addition is not nearly so dramatic.

\end{document}